\documentclass[letterpaper,11pt,journal]{IEEEtran}
\usepackage[T1]{fontenc}
\usepackage{amsmath,amssymb}
\usepackage{graphicx}
\usepackage{array,booktabs,tabularx,multirow,cite,xcolor}
\usepackage{tabularray}
\UseTblrLibrary{booktabs}

\usepackage[hidelinks]{hyperref}
\usepackage{microtype}
\newcolumntype{Y}{>{\raggedright\arraybackslash}X}
\newcolumntype{P}[1]{>{\raggedright\arraybackslash}p{#1}}
\graphicspath{{资料/figures/embodied_multiagent_perception_20261022/}{figure/}}
\begin{document}

\title{Embodied AI for Radio Mapping: From Solo to Swarm}
\author{Shangjie~Zhuang, Wenlihan~Lu, Zhentian~Zhang, Ziwei~Huang, Zaichen~Zhang, Rongqing~Zhang, and Shijian~Gao%
}
\maketitle

\begin{figure*}[!t]
    \centering
    \includegraphics[width=\textwidth]{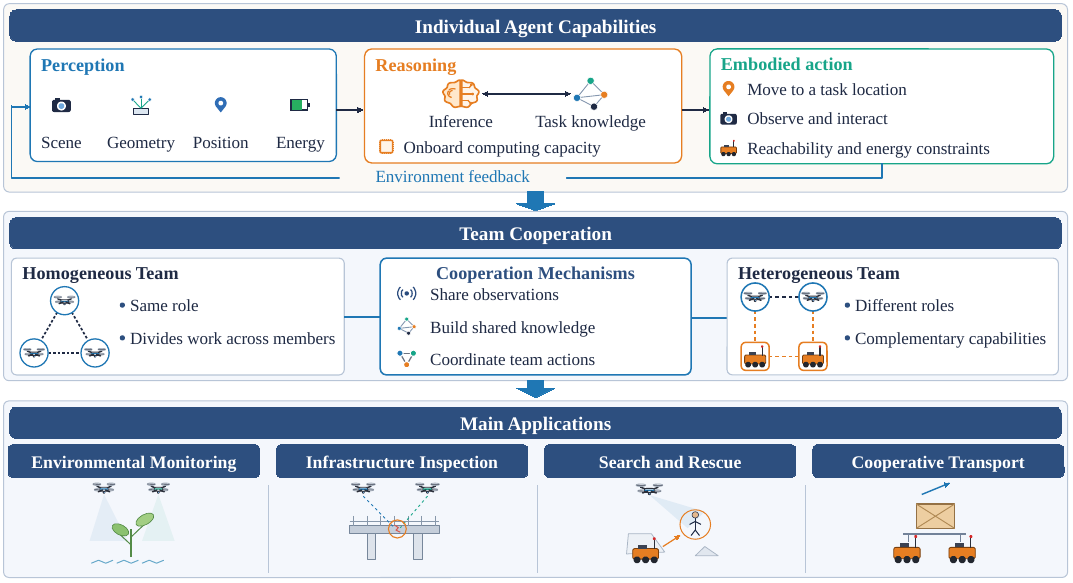}
    \caption{Embodied multi-agent cooperation and its applications.}
    \label{fig:task-expansion}
\end{figure*}

\begin{abstract}
Radio maps provide spatial knowledge of wireless environments to assess coverage, manage spectrum use, and support connectivity-aware mobility. Recent advances in embodied intelligence (EI) offer a promising approach to radio map reconstruction by enabling agents to perceive the environment and autonomously adjust their survey paths toward safe, reachable locations. However, a single agent remains constrained by limited endurance and onboard computing capacity, limiting measurement collection and reconstruction performance. Swarm cooperation can support more efficient radio mapping over larger areas by sharing observations and coordinating members' sampling. This article first explores the capabilities and applications of cooperative embodied agents. A cooperative radio map reconstruction framework is then presented, together with its key enabling techniques. Case study results illustrate how scene perception supports access to informative sampling locations and how air--ground cooperation improves reconstruction through complementary measurements and mobile support. The article concludes with open challenges and future research directions for reliable and efficient embodied cooperative radio mapping in complex environments.
\end{abstract}

\section{Introduction}\label{sec:intro}
Radio maps characterize the spatial distribution of wireless signal characteristics, providing radio environment knowledge for network operation and autonomous systems~\cite{romero2022}. They help identify coverage gaps for network deployment, reveal interference patterns for spectrum management, and support connectivity-aware route planning for mobile robots and uncrewed aerial vehicles (UAVs). These applications motivate accurate radio map reconstruction across diverse environments, including low-altitude airspace where aerial services require reliable wireless connectivity~\cite{3gpp22829}. Preset-route surveying acquires measurements along constrained trajectories for reconstructing three-dimensional radio maps~\cite{hu2023}. However, reconstructing these maps in complex, dynamic, and unknown environments is challenging, as agents must perform real-time navigation under strict onboard resource constraints.

To meet these demands, embodied intelligence (EI) enables agents to perceive their surroundings and reason about tasks to guide physical action in partially known environments. Vision-and-language navigation for UAVs illustrates how scene observations and instructions guide physical movement~\cite{liu2023aerialvln}. In radio mapping, agents actively select accessible waypoints, collect informative measurements, and reconstruct radio maps~\cite{shrestha2023}. Active perception in uncharted environments further incorporates partial geometry into waypoint selection under a travel budget~\cite{lu2026}. Nevertheless, a single agent remains constrained by its accessible waypoints, flight endurance, and onboard computing capacity.

Multi-agent cooperation can alleviate these limitations by distributing survey tasks and combining observations. Teams can be categorized by their roles: members of a homogeneous team perform the same duties, such as sensing. Existing multi-UAV studies optimize sampling trajectories with diffusion policies~\cite{zhao2026temporal} or jointly learn radio mapping and navigation~\cite{wang2026crmn}. Beyond parallel sensing, agents may differ in their reachable locations, sensors, and onboard resources~\cite{cladera2024mocha}. A heterogeneous team combines agents with different duties. For example, members can undertake sensing or support duties according to their onboard resources and physical capabilities. Highly mobile sensing agents focus on surveying and sampling, while support agents contribute their own measurements and provide additional processing capacity. Fig.~\ref{fig:task-expansion} summarizes these cooperative capabilities and their broader applications. 

Applying these complementary capabilities to radio map reconstruction requires matching the task requirements with members' capabilities. Conventional rule-based or optimization-based planning typically starts from explicitly defined tasks and constraints, requiring human instructions and scene observations to be translated into suitable planning inputs. Multimodal large language models (MLLMs) have attracted interest because they integrate instructions and visual information for embodied task planning~\cite{driess2023palme}. For radio mapping, MLLMs can combine instructions with scene images to interpret survey objectives and identify target regions. Pretraining can help models generalize to new survey tasks and unfamiliar scenes, reducing reliance on mission-specific workflows. Retrieval-augmented generation retrieves members' capability descriptions. The MLLM then considers these descriptions alongside members' current states to assign tasks.

This article reviews the transition from individual embodied surveying to cooperative radio map reconstruction, examining how shared observations and complementary roles extend individual mapping capabilities. MLLM-assisted planning matches survey requirements to members' capabilities, clarifying their duties to improve reconstruction performance. The main contributions are summarized as follows:
\begin{itemize}
    \item We explore how shared observations and complementary support extend individual embodied sensing, and present a cooperative radio mapping framework that matches tasks to members' capabilities and enables cooperative mapping through information exchange.
    \item We examine the proposed framework's key enabling techniques: MLLMs interpret requirements and assign duties according to members' capabilities, active sampling selects subsequent measurement locations, and information sharing and collaborative inference support information exchange and rapid decisions.
    \item We identify open challenges in task planning, embodied sensing, and cooperative support, and outline research directions for reliable and efficient cooperative radio map reconstruction.
\end{itemize}

\section{An Overview of Embodied Multi-Agent Cooperative Radio Mapping}\label{sec:multiagent}
In this section, we first examine how swarm cooperation extends solo surveying, then introduce a framework that organizes members' capabilities for radio mapping.

\subsection{Cooperative Capabilities of Embodied Agents}
An embodied agent adapts its survey based on its own observations. From its map and measurements, it selects accessible sensing locations, collects radio-frequency (RF) observations, and estimates signal strength in unmeasured regions. Visual or geometric observations help identify obstacles and feasible routes, directing sensing toward regions requiring further measurements. Nevertheless, limited physical access restricts coverage, while onboard resources constrain endurance and reconstruction.

A team of embodied agents can survey different regions in parallel, extending coverage over larger areas within a limited mission time. Fusing their observations incorporates measurements collected beyond each member's own survey into the shared map. The value of sampling a location also depends on what teammates have measured or plan to measure. Shared map updates and intended visits help members avoid unnecessary overlap and prioritize locations where further measurements are expected to improve reconstruction. Members can measure at locations beyond teammates' reach. Shared scene observations help members lacking a particular sensor assess reachable sampling locations. Beyond sharing observations, members can take on support duties according to their computing resources and connectivity. These support agents process or relay measurements and map updates alongside their own sensing, allowing teammates to devote more onboard resources to surveying. AeroDuo illustrates complementary roles by assigning scene-level reasoning and local navigation to UAVs at different altitudes~\cite{wu2025aeroduo}. Capability-based assignments~\cite{chen2025emos} integrate these sensing and support duties for collective reconstruction. Air--ground teams provide one example, combining aerial sensing with ground-based sensing and support.

With swarm cooperation, each agent can contribute to a radio map that extends beyond what it could observe or reconstruct alone. The following framework describes how agents with different roles work together to achieve this shared reconstruction objective.

\subsection{Framework for Cooperative Radio Mapping}
\begin{figure*}[!t]
    \centering
    \includegraphics[width=\textwidth]{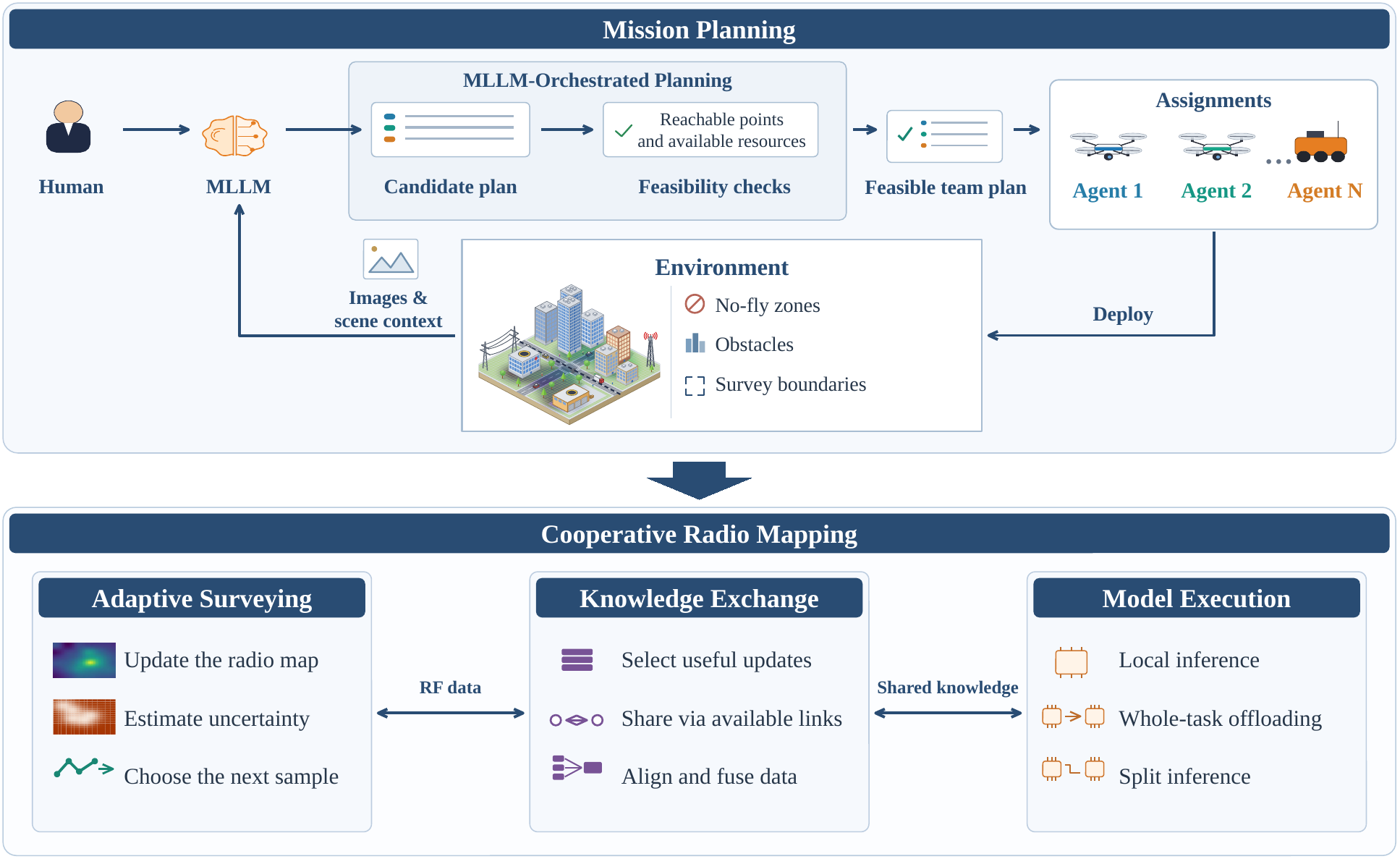}
    \caption{A framework for cooperative radio mapping with embodied agents.}
    \label{fig:framework}
\end{figure*}

Fig.~\ref{fig:framework} illustrates the two-layer organization of the proposed framework. In the upper layer, human interaction with an MLLM translates the mapping objective into team tasks that undergo feasibility checks before assignment to individual members. The lower layer organizes Adaptive Surveying, Knowledge Exchange, and Model Execution around members' sampling in the physical environment. These functions exchange measurements, map updates, and inference requests and results, allowing members to adjust subsequent sampling as new information becomes available. The functions can run on the same member or be distributed across members according to their capabilities and assignments.

\textbf{Mission Planning:} The team planner uses an MLLM to combine human instructions with scene information, identify survey regions and measurement needs, and generate candidate assignments~\cite{driess2023palme}. The planner checks whether members can reach the assigned locations and acquire the required measurements within their energy and computing budgets, then revises infeasible assignments. The final plan defines each member's duties and scope for autonomous adjustments. The planner can run on a support member with sufficient computing resources, while agents execute autonomously within their assigned scope.

\textbf{Adaptive Surveying:} Members select sampling locations and routes according to their assigned regions and priorities. Scene observations help identify obstacles and reachable locations, while the current map, predictive uncertainty, and measurement history indicate where further sampling is worthwhile. Vision--language--action (VLA) models can translate scene observations and instructions into actions~\cite{kim2025openvla}, while geometry supports route planning and obstacle avoidance. New radio measurements update the map and its uncertainty, and the revised estimates guide subsequent sampling. Sharing teammates' measurements and plans also allows members to account for others' sampling contributions and reduce unnecessary duplication.

\textbf{Knowledge Exchange:} Members select measurements and map updates for sharing and adjust transmission timing according to cooperative reconstruction needs and available links. Receivers align their own and teammates' observations by acquisition location and time, accounting for differences in measurement conditions during fusion to update shared knowledge. Exchanging map estimates and survey progress allows members to use information about regions they have not yet visited to adjust subsequent actions. During disconnection, members retain unsent measurements for later sharing when a link becomes available and the information remains relevant.

\textbf{Model Execution:} This function determines which members execute the inference tasks needed for map reconstruction and sampling decisions, and how computation is distributed among them. Members can complete a task locally or offload the whole task to a teammate. For models that support staged execution, a member can run the initial stages locally and let a teammate complete the remaining computation. The execution choice accounts for computing load and connectivity, including the time and energy required for transmission, waiting, and processing. Support members can contribute their own measurements while undertaking additional computation, freeing other members' resources for surveying. Returned inference results update the map or guide sampling adjustments.

\section{Key Enabling Techniques}\label{sec:techniques}
This section explains representative mechanisms for implementing the proposed framework, from initial task planning to cooperative sampling, information sharing, and computation.

\subsection{MLLM-Assisted Team Planning}
Task decomposition and capability matching translate mission requirements into executable survey assignments. An MLLM relates visual scene information to instructions and member states to identify survey regions, required altitudes and bands, priorities, and restricted areas~\cite{driess2023palme}. Robot specifications and kinematic tools help describe members' physical capabilities~\cite{chen2025emos}.

Using these descriptions and mission requirements, the MLLM generates candidate sensing and support assignments. Using geometry and resource checks, the planner verifies that assigned members can reach target locations and measure required bands. Camera or light detection and ranging (LiDAR) observations support reachability checks. It revises infeasible assignments and compares feasible plans by completion time and remaining energy. The initial plan assigns duties and defines the scope of members' autonomous adjustments. Members refine sampling locations as new measurements arrive.

\subsection{Active Sampling and Radio Map Reconstruction}
Probabilistic mapping exploits spatial and spectral correlations to update signal estimates and predictive uncertainty~\cite{shrestha2023}. The estimate at a queried position, frequency, and time is expressed as
\begin{equation}\label{eq:cooperative_reconstruction}
\widehat M(\mathbf x,f,t)=\mathcal R(\mathbf x,f,t\mid\mathcal D_t),
\end{equation}
where $\mathbf x$, $f$, and $t$ denote position, frequency, and time. $\mathcal D_t$ indicates the available measurements, including new and retained historical measurements from sensing and support members. Acquisition times are retained to assess temporal relevance.

Information-driven planning selects feasible team actions according to estimated benefit:
\begin{equation}\label{eq:cooperative_sampling}
 a^{\star}\in\underset{a\in\mathcal A}{\arg\max}\;G(a)
\end{equation}
Here, $\mathcal A$ contains team sampling actions permitted by physical and resource constraints, and $G$ estimates the expected reduction in reconstruction uncertainty.

One implementation uses receding-horizon informative path planning: agents generate candidate sampling points and short routes, estimate their gains, and select near-term actions within movement and energy budgets. Members select routes in sequence, evaluating remaining candidates for additional measurements beyond teammates' planned visits. These comparisons account for overlap between planned measurements. Members begin their routes and use new measurements to update the map and subsequent plans. Active verification checks potentially misleading sampling guidance. Agents compare RF measurements with predictions saved before sampling, accounting for receiver noise and positioning errors. Persistent discrepancies raise the priority of further checks, including in regions of low predicted uncertainty.

\subsection{Adaptive Information Sharing and Fusion}
Members adapt what they share and when they transmit to teammates' mapping needs and available links. Event-triggered sharing sends updates when map estimates change substantially or revised plans affect teammates' upcoming work.

During a limited contact, data selection can be summarized as
\begin{equation}\label{eq:selective_sharing}
\mathcal S^{\star}\in\underset{\operatorname{size}(\mathcal S)\le B}{\arg\max}\;V(\mathcal S)
\end{equation}
Here, $\mathcal S$ contains the measurements and map updates chosen for transmission. $B$ is the number of bits that can be sent during the contact. $V(\mathcal S)$ estimates how much these data would improve the receiver's map. The sender first identifies unshared measurements and map updates using the receiver's shared records. It then uses the receiver's map summary to estimate the additional improvement from each item. A greedy implementation selects the item with the largest estimated improvement per bit that fits the remaining capacity, then reassesses the other items to account for overlapping information. As link quality and expected contact duration change, members reestimate how much they can send and adjust the transmission order. Unsent measurements and map updates are saved for later contact or relayed through suitable peers using gossip-based exchange~\cite{cladera2024mocha}. Members balance forwarding costs with their own sampling needs.

The receiver aligns measurements by acquisition location, frequency, and time, accounting for differences in receiver calibration. Unique measurement identifiers prevent duplicate use of the same observation. For updates containing only changes to an earlier map, the receiver checks that the reference version matches its stored map. Updated estimates then inform subsequent sampling.

\subsection{Collaborative Model Inference}
Collaborative inference lets members share the processing needed for map reconstruction and sampling decisions. An agent can complete a task locally or offload the whole task to a teammate. It estimates the time and energy required by each option, accounting for data transfer and waiting for remote processing. Offloading is useful when faster processing compensates for these delays and the result arrives in time to guide sampling.

Models that support staged execution offer another option: split inference. A sensing member runs the initial stages and sends their outputs to a support member, which completes the remaining stages. The agent evaluates possible split points by estimating local and remote processing times and the time to transmit intermediate outputs. An early split can save onboard computation, but its intermediate outputs may require more transmission than the original input. For several jobs, choosing split points together with an execution schedule helps coordinate processing and transmission~\cite{duan2024offloading}.

When several requests compete for the same support member, it prioritizes results needed for upcoming sampling decisions and delays or redirects other requests. These arrangements adapt to computing load and connectivity while accounting for each member's own sensing tasks. While waiting for a shared result, an agent continues sampling with its current map and local observations, then revises its plans when the result arrives.

\section{Case Studies: From Single-Agent to Air--Ground Cooperative Radio Mapping}\label{sec:cases}
This section examines how map feedback and team configurations affect radio map reconstruction, and how scene perception supports access to informative sampling locations. A low-altitude air--ground survey provides the case study.

\begin{figure*}[!t]
    \centering
    \includegraphics[width=\textwidth]{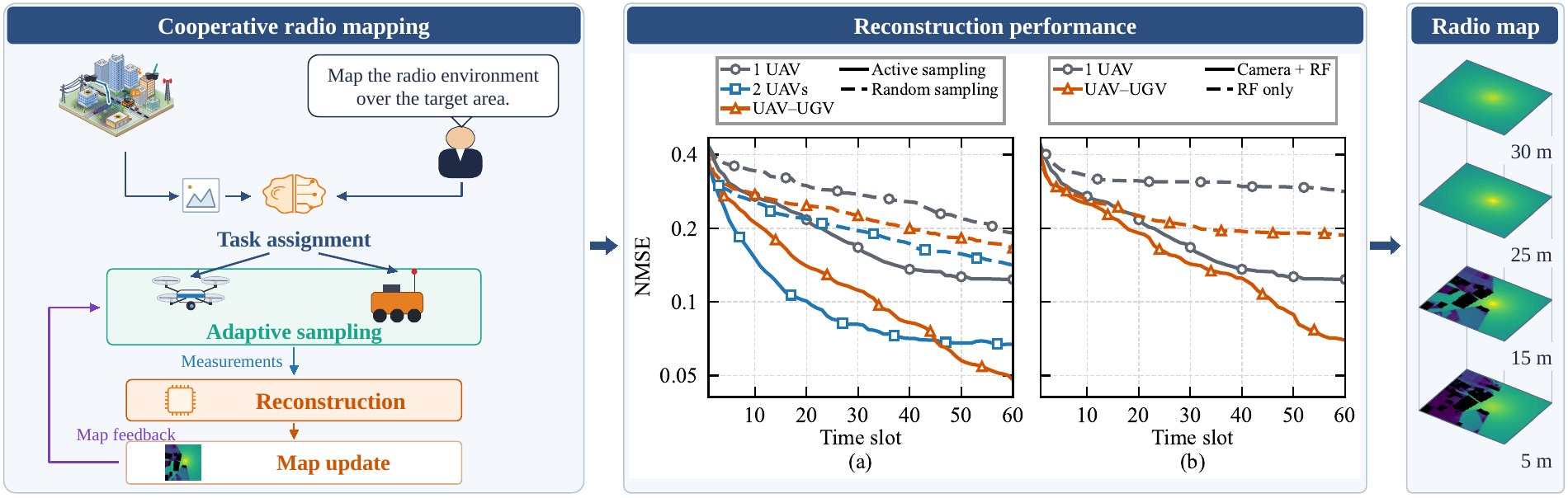}
    \caption{Radio map reconstruction under different sampling strategies and agent configurations.}
    \label{fig:nmse}
\end{figure*}

\subsection{Experimental Setup}
We evaluate radio map reconstruction using the FARM dataset~\cite{gao2026farm} and measure accuracy by the normalized mean squared error (NMSE). Table~\ref{tab:setup} summarizes the main numerical settings.

\begin{table}[!t]
\centering
\caption{Case study settings.}
\label{tab:setup}
\small
\renewcommand{\arraystretch}{1.12}
\setlength{\tabcolsep}{4pt}
\begin{tabularx}{\columnwidth}{@{}P{0.60\columnwidth}Y@{}}
\toprule
\textbf{Parameter} & \textbf{Value} \\
\midrule
Survey area & 200~m $\times$ 200~m \\
Horizontal grid cell size & 2~m $\times$ 2~m \\
Map altitude layers & [5, 15, 25, 30]~m \\
UAV flight altitude & 15--30~m \\
UGV sampling altitude & 5~m \\
Link SNR threshold & 0~dB \\
\bottomrule
\end{tabularx}
\end{table}

The building layout is not known in advance, so agents use local scene observations to identify reachable sampling locations. Aerial agents carry downward cameras whose footprints grow with altitude, while the ground agent observes its surroundings within 20~m. These observations define the reachable candidate waypoints. The UAV sensing policy, trained with proximal policy optimization (PPO), then selects the next waypoint using the current map estimate, its uncertainty, and the UAV's own state. To reach the selected waypoint and collect radio measurements, the agent plans a route through the free space it has observed. For reconstruction, aerial measurements are sent over an air--ground link whose quality depends on distance and building blockage. When the link is unavailable, measurements are saved until transmission becomes possible. An ensemble network reconstructs the map from the delivered measurements, and the updated map guides subsequent sampling.

\textit{Solo and homogeneous teams:} One UAV, or a homogeneous team of two UAVs with the same sensing role, samples the 15--30~m layers and reports to a fixed base station near the center of the area, which reconstructs and returns the map without sensing.

\textit{Heterogeneous team:} An uncrewed ground vehicle (UGV) joins one UAV as the support agent. It samples the 5~m altitude layer, hosts the reconstructor, and returns the map to its aerial teammate, moving under a heuristic that keeps the link available.

We compare active sampling, in which the agents select waypoints using the current map estimate, with random sampling that ignores the map. An RF-only variant removes the cameras, so the agents do not know the building layout and an agent that hits a building stops its mission.

\subsection{Results and Discussion}

Fig.~\ref{fig:nmse}(a) shows that active sampling achieves a lower final NMSE than random sampling across all three configurations, illustrating the value of using map feedback to guide measurement collection. The homogeneous two-UAV team reduces NMSE fastest initially through parallel sampling of the aerial layers, but its improvement later slows, reaching 0.067. The solo agent and the homogeneous team frequently lose their links to the fixed base station. The UAV--UGV team continues improving and reaches 0.048. Its ground member contributes measurements from the 5~m layer and moves to support the delivery of aerial measurements to the reconstructor. The comparison illustrates how complementary sampling and mobile support can improve reconstruction beyond the initial gains from parallel aerial surveying.

Fig.~\ref{fig:nmse}(b) examines the effect of scene perception. Without cameras, the agents select waypoints and routes that cross unseen buildings, and many missions end in collisions that stop sampling early. The final NMSE rises for the solo case and the heterogeneous team. With cameras, each agent identifies reachable locations in its view and plans collision-free routes to collect further measurements. Scene perception therefore helps agents turn sampling decisions into acquired RF observations, supporting reconstruction in obstructed environments.

\section{Challenges}\label{sec:challenges}
The proposed cooperative radio mapping framework needs to translate mission requirements into member assignments, use spatial observations to guide dynamic sampling, and support map reconstruction through information sharing and collaborative inference. These functions are interdependent, and their coordinated implementation faces the following challenges.

\subsection{Reliability of MLLM-Assisted Team Planning}
An MLLM combines instructions with physical observations to assign duties~\cite{driess2023palme}, but decomposition may omit required altitudes or assign bands beyond a member's sensing capability. Individually feasible tasks may also require the same member simultaneously. Such errors can leave regions without required measurements, degrade map reconstruction quality, or even prevent the team from completing the required reconstruction task. Feasibility checks only assess the task requirements explicitly specified during planning. If a requirement is omitted, a plan may pass these checks yet fail to fulfill the original mission. Reliability therefore requires both executable assignments and a complete, accurate translation of requirements into tasks and constraints.

\subsection{Multimodal Spatial Understanding under Partial Observations}
Executing team assignments requires correct reachability judgments of sampling locations. Members may need observations from sensors they lack. Using these observations requires identifying the same physical locations or structures across different sensing modalities and viewpoints. However, partially overlapping observations may provide too few common features for reliable matching. Positioning errors further complicate matching when similar structures occur at different locations. Such mismatches can make blocked passages appear traversable or reconstruct a wrong map. The difficulty is determining whether shared observations sufficiently support local spatial understanding and how alignment errors affect reachability and measurement localization.

\subsection{Dynamic Cooperative Sampling}
Spatial understanding identifies feasible sampling locations, while dynamic sampling determines which are worth visiting. Equation~\eqref{eq:cooperative_sampling} ranks feasible team actions by expected uncertainty reduction~\cite{shrestha2023}. As teammates collect new measurements or revise their routes, some planned samples may duplicate information already available or expected from others, making alternative sampling locations more useful for reconstruction. However, wrong estimated uncertainty can misguide sampling by causing agents to repeatedly skip regions that still require verification, leaving reconstruction errors uncorrected. Active verification is therefore needed to check predictions that may misguide sampling. When new measurements differ substantially from prior predictions, agents may need to switch to another sampling criterion. The main difficulties are determining when to switch criteria and how to select suitable alternative metric that can guide subsequent sampling toward improving reconstruction.

\subsection{Selective Sharing over Dynamic Connections}
During surveying, members exchange observations to update maps using teammates' measurements and adjust subsequent sampling locations and routes. Equation~\eqref{eq:selective_sharing} selects updates by expected receiver benefit, but the sender may not know the receiver's latest measurements or incoming updates. Movement and blockage can shorten contacts or remove relay paths before transmission completes. Meanwhile, new observations at the receiver can provide information it already has, reducing the contribution to map reconstruction. Learning the receiver's state can reduce redundant sharing but consumes capacity needed for data delivery. During brief contacts, excessive state exchange can leave too little time for useful updates. Assessing information value and delivering data thus compete for the same contact capacity and duration.

\subsection{Reliable and Timely Collaborative Inference}
Inference on shared observations must finish in time to improve reconstruction and guide sampling. When a model is split across members, later stages await preceding computation and delivery of intermediate results. Even members with spare computing capacity may remain idle awaiting inputs. When requests share a processing member, the fastest partition for one task can increase others' waiting, so local acceleration may not improve overall completion time~\cite{duan2024offloading}. Movement and workload changes alter transmission and processing times. Reassigning computation may require transferring intermediate results or repeating earlier stages. Timeliness depends on when results support specific actions: a late result may still refine the map but miss its intended sampling decision.

\section{Future Directions}\label{sec:future}
Building on the preceding challenges, we examine future directions for cooperative radio mapping from the perspectives of verifiable team planning, spatial understaing, and connectivity prediction and knowledge transfer under limited communication.

\subsection{Verifiable MLLM-Assisted Team Planning}
Future research could combine multimodal task interpretation with constraint solving to develop verifiable team plans. An MLLM could translate mission instructions and scene observations into explicit task requirements, which can be checked against their sources for omissions or misinterpretations. Based on these requirements and member capabilities, a constraint solver could generate assignments that account for dependencies between members' tasks. Conflicting constraints would be returned to the MLLM for revision, while feasible plans could be compared by their expected reconstruction benefit within the mission budget. Open directions include how uncertainty in scene information and member capabilities should affect plan acceptance, and how checking and revision time grows with team size. Evaluation should examine whether this process reduces planning errors and improves reconstruction without introducing excessive planning delays.

\subsection{Physics-Informed Spatial Intelligence}
A UAV swarm could assign complementary scene observation and RF sampling tasks according to members' positions, reachable altitudes, and sensor configurations to jointly build a propagation model. Members could observe structures hidden from teammates' views, while RF measurements from different locations constrain and calibrate propagation predictions for radio map reconstruction. Research could examine whether joint observations reduce model ambiguities left by individual members' local data under incomplete transmitter information and imperfect scene understanding. For example, one member could add observations of building geometry while teammates collect RF measurements at relevant locations, jointly testing whether the resulting propagation model explains their measurements. A key question is which combinations of members and observations best support model refinement.

\subsection{Predictive Sensing--Communication Planning}
Radio world models could combine link observations with scene geometry to predict connection opportunities along planned routes. An aerial agent that predicts its link to the ground support agent will last only a few seconds after passing the next building could send its most informative measurements first, or keep sampling if a longer contact follows. Research should establish how far ahead such predictions remain useful and whether they improve reconstruction under matched time and energy budgets.

\subsection{Cross-Model Knowledge Transfer}
Members could share knowledge rather than full models. Through knowledge distillation, the large reconstructor on the ground support agent could transfer its predictions for the surveyed area to compact aerial models. Aerial agents could then use the updated models to guide sampling and return measurements for further learning. Research should determine which knowledge is worth transferring over limited links, how local measurements can test received predictions, and how to update models without retraining them in full.

\section{Conclusion}
This article reviewed the transition from solo surveying to swarm reconstruction and presented a framework for cooperative radio mapping with embodied agents. We discussed how MLLM-assisted planning assigns team duties, while active sampling, adaptive information sharing, and collaborative inference use shared knowledge to support reconstruction. The case study showed that scene perception enables safe sampling, while ground observations and mobile support jointly improve reconstruction. Finally, we outlined future research directions for radio map reconstruction in complex wireless environments, focusing on planning reliability, the integration of spatial understanding with propagation modeling, and cooperative efficiency.


\end{document}